# Influence of microstructural and crystallographic inhomogeneity on tensile anisotropy in thick-section Al-Li-Cu-Mg plates

**Authors:** X. Xu[a, *], M. Hao[b, *], J. Chen[b], W. He[b], G. Li[b], C. Jiao[c], T.L. Burnett[a, d, **], X. Zhou[a, **]

[a] Department of Materials, University of Manchester, Manchester, M13 9PL, UK
[b] Beijing Institute of Aeronautical Materials, Beijing, 100095, China
[c] Thermo Fisher Scientific, Achtseweg Noord 5, 5651 GG, Eindhoven, the Netherlands.
[d] Henry Royce Institute for Advanced Materials, The University of Manchester, Manchester, M13 9PL, UK

* These authors contributed equally to this work.

** Corresponding author:
timothy.burnett@manchester.ac.uk
xiaorong.zhou@manchester.ac.uk

## Abstract

Thick-section plates made from a recently developed Al-Cu-Mg-Li alloy have been evaluated to understand the influence of microstructure on the anisotropy of tensile strengths after natural and artificial ageing treatment. Pancake-shaped grains with a coarse substructure and strong crystallographic texture with a β-fibre orientation at the mid-thickness position are observed. In addition, an inhomogeneous distribution of $T_1$ precipitates through the plate thickness has been revealed with the volume fraction of intragranular precipitates ~ 40% higher at the plate centre than the ¼ thickness position. Altogether these microstructural features contribute to the in-plane anisotropy of tensile strengths that is ~ 5% higher at the mid-thickness position than the ¼ thickness position. The variation of ageing-induced $T_1$ precipitates through the plate thickness further contributes to the through-thickness anisotropy that is ~ 3% higher in T8 temper as compared to T3 temper.

# 1. Introduction

Aluminium-lithium (Al-Li) alloys have been used in a wide range of applications in the aerospace industry due to their excellent specific strength and good fatigue performance [1]. They have been specified for use in the structural components including fuselage, bulkhead and wing structures bringing a higher fuel efficiency and improved performance to modern aircrafts [2]. After decades of development, the modern Al-Li alloys also contain Cu, Mg, Mn, Zn and Zr as alloying elements, with Li in the range of 1 - 2 wt. % [1,2]. The alloys are normally supplied in product forms such as sheets and plates following a combination of thermomechanical processing of rolling, solid solution treatment and ageing [1,3]. The final microstructure arising from thermomechanical treatment typically comprises elongated, pancake-shaped grains in combination with constituent intermetallic phases, dispersoids and precipitates [1,4–6].

The anisotropy of mechanical properties is well-known for components made from Al-Li alloys. This typically involves the in-plane anisotropy of the properties in the L-LT plane and the through-thickness anisotropy associated with the variation through the plate thickness [3,7]. Previous studies have highlighted a variety of microstructural and crystallographic factors related to the in-plane anisotropy of mechanical properties for Al-Li alloys, including crystallographic texture, grain morphology and the characteristics of second phases [1,3,8–10]. For instance, strong crystallographic texture with a β-fibre orientation causes a high level of in-plane anisotropy of tensile strength and ductility in 2195-T8 plates [9]. The pancake-shaped grain morphology of an unrecrystallised microstructure is also a factor in the in-plane anisotropy of fracture toughness in an 8090-T8 plate [10]. In addition, the uneven distribution of $T_1$ ($Al_2CuLi$) precipitates on the {111} plane variants increases the in-plane anisotropy in the 2090 and 2198 alloys [11,12].

The development of alloy composition and thermomechanical processing has led to a reduction of in-plane anisotropy in Al-Li products. For instance, the modern third-generation Al-Li alloys have balanced contents of Cu, Mg and Li with the addition of Mn [1,3]. This introduces dispersoid and precipitate phases that help reduce crystallographic texture anisotropy after thermomechanical processing [1,3,8,13]. The fabrication route as described in [3,14] also introduces an intermediate annealing process in a multi-stage rolling practice to decrease in-plane anisotropy by reducing crystallographic texture. In addition, the emerging techniques such as asymmetric rolling and snake rolling has demonstrated the ability to produce Al-Li plates with a reduced level of in-plane anisotropy [15,16].

However, for the production of thick plate Al-Li materials it is a further challenge as the through-thickness anisotropy has to be controlled as well and the ability to tightly control the processing conditions becomes more limited. The mechanical and corrosion property anisotropy continue to limit the application of thick-section Al-Li alloys due to the significant crystallographic and microstructural inhomogeneities that can emerge. For instance, the tensile elongation at the mid-thickness position is ~ 20% lower than the material at the surface in a 90-mm thick Al-Cu-Li (T84) plate due to strong crystallographic texture with the β-fibre orientation and pancake-shaped grain morphology [17]. In addition, a 95-mm thick 2297-T87 plate exhibits the highest susceptibility to localised corrosion at the quarter-thickness position with the highest volume fraction of coarse constituent intermetallic and precipitate phases [18].

The through-thickness heterogeneities of microstructure and crystallographic texture are usually caused by variation of deformation strain and temperature (i.e. cooling rates) in thick-section components during thermomechanical processing. For instance, the variation of deformation strain from hot rolling caused an evident gradient of crystallographic texture through the plate thickness with well-developed β-fibre texture components at the centre of a 2195-T3 plate [19]. The lower cooling rate at the mid-thickness position also resulted in the formation of large, quench-induced $T_1$ phases in a thick-section plate [20]. In addition, the distribution of coarse constituent intermetallic and precipitate phases varies from surface to centre due to the variation of deformation strain in a 2297-T87 plate [18]. These studies highlight the difficulty in obtaining a uniform microstructural and crystallographic condition for reducing the anisotropy of mechanical properties in thick-section plates.

Here we investigate the microstructure and mechanical properties of newly developed thick-section plates of Al-Li-Cu-Mg alloy at different tempers. The examinations of microstructure were conducted using multiscale electron microscopy based techniques to obtain the quantitative data of grain structure and second phase particles. The influence of microstructural heterogeneities on mechanical performance is further discussed based on the results of uniaxial tensile testing. The outcomes of the present work are intended to provide critical information for improvement of alloy design and thermomechanical processing that will minimise mechanical anisotropy in thick-section components.

## 2. Material and experimental procedure

*2.1 Material*

The composition specification of the alloy is given in Table 1. The as-received material had been hot-rolled to a final thickness of 50 mm and then solution treated at ~ 520 °C, followed by water quenching to room temperature. The plate was pre-stretched at room temperature to a strain of ~ 4%, followed by a natural ageing process for over three months to obtain T3 condition. For the T8 condition the identical procedure of rolling, solution heat treatment and pre-stretch was followed by a two-stage artificial ageing process at 120 °C for 20 hours and then 145 °C for 12 hours to reach the peak-aged condition.

**Table 1. The nominal composition (wt. %) of the Al-Li-Cu-Mg alloy investigated in this study.**

| Al | Cu | Li | Mg | Mn |
|---|---|---|---|---|
| bal. | 3.0 – 4.0 | 1.2 – 1.7 | 0.2 – 0.6 | 0.2 – 0.6 |
| Fe | Zr | Si | Zn | |
| ≤ 0.1 | 0.1 – 0.2 | ≤ 0.10 | 0.2 – 0.8 | |

*2.2 Uniaxial tensile testing*

To examine the mechanical anisotropy, flat dog-bone tensile specimens were machined from the L–LT and the LT–ST planes at the ¼ thickness (T/4) and mid-thickness (T/2) positions for both T3 and T8 tempers, as illustrated in Figure 1. In line with the gauge dimension as specified in GB/T 228.1–2010 [21], the specimens along the L and the LT directions measure 25 mm in gauge length and 5 mm in gauge diameter, whilst the specimens along the ST direction measure 15 mm in gauge length and 3 mm in gauge diameter. The specimens were tested under uniaxial tension following the testing procedure as described in the same standard [21]. The tensile testing was carried out at room temperature and at a strain rate of 0.04 min$^{-1}$ prior to the 0.2% offset then 0.2 min$^{-1}$ to failure. Two tests were conducted to obtain the average values of tensile strengths and elongation in both tempers and thickness positions.

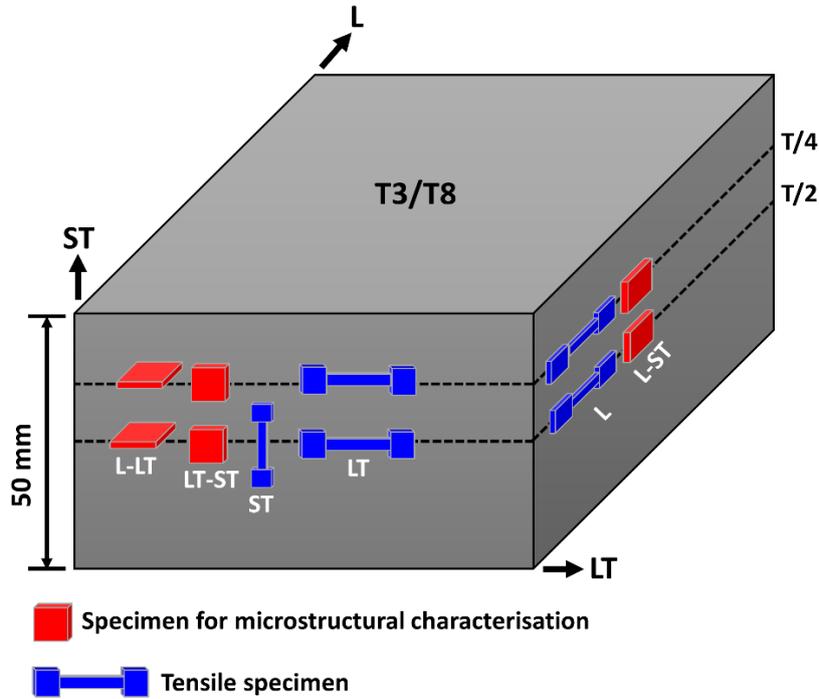

**Figure 1. A schematic diagram illustrating the specimens extracted from rolled plate for microstructural characterisation and uniaxial tensile testing.**

The in-plane anisotropy (IPA) was initially assessed based on the difference of tensile strengths following the formula [22,23]:

$$IPA = \frac{(N-1)\sigma_{max} - \sigma_{mid1} - \sigma_{mid2} - \cdots - \sigma_{min}}{(N-1)\sigma_{max}} \times 100\% \qquad (1)$$

Where N is the number of stress loading directions, $\sigma_{max}$, $\sigma_{min}$ and $\sigma_{mid}$ are the maximum, minimum and intermediate values of tensile strengths measured within the L-LT plane, respectively. In this study, two values were measured from the L and the LT directions at the same thickness positions. Therefore, the in-plane anisotropy was expressed as:

$$IPA = \frac{\sigma_L - \sigma_{LT}}{\sigma_L} \times 100\% \qquad (2)$$

Where $\sigma_L$ and $\sigma_{LT}$ are the tensile strengths measured from the L and the LT directions, respectively.

The through-thickness anisotropy (TTA) was further evaluated based on the same principle using the equation:

$$TTA = \frac{\sigma_{T/2} - \sigma_{T/4}}{\sigma_{T/2}} \times 100\% \quad (3)$$

Where $\sigma_{T/2}$ and $\sigma_{T/4}$ are the tensile strengths measured from the T/2 and the T/4 positions, respectively.

*2.3 Quantitative microstructural evaluation*

The specimens for microstructural characterisation were extracted from L-LT, L-ST and LT-ST planes at the T/4 and T/2 positions for both T3 and T8 tempers, as illustrated in Figure 1. The specimens were 10 × 10 mm and 3 mm in thickness. These specimens were prepared by grinding with SiC grinding papers using water as lubricant, polishing with 3 μm and 1 μm diamond suspensions and a final chemo-mechanical polishing in a solution of 1:1.5 amorphous colloidal silica suspension (Buehler Mastermet, 0.06 μm) to deionised water to give a surface finish suitable for EBSD analysis.

Electron backscatter diffraction (EBSD) mapping was carried out using a Tescan Mira 3 field emission gun scanning electron microscope (FEG-SEM) equipped with an Oxford Instruments Symmetry CMOS acquisition system at an accelerating voltage of 20 kV and a step size of 2 μm to provide an overview of grain structure from the L, the LT and the ST directions. Each map covered an area of 4 × 4 mm and was composed of over 4,000,000 data points collected from over 1,000 grains within 2 hours. The Kikuchi patterns were collected at a pixel resolution of 156 × 128, a beam current of ~ 10 nA and an exposure time of ~ 1 ms with an indexing rate > 80% confidence.

In addition, high resolution EBSD mapping was conducted over the regions measuring 20 × 20 μm at a step size of 0.1 μm to reveal the details of grain structure. The analysis was performed at an accelerating voltage of 10 kV to reduce the interaction volume. The Kikuchi patterns were collected at a pixel resolution of 156 × 128, a beam current of ~ 7 nA and an exposure time of ~ 10 ms with an indexing rate > 95% confidence.

The post-processing of the EBSD data was conducted using the HKL Channel 5 software version 5.12. The dimension of grains as identified with a boundary misorientation of > 15° was measured along the L, the LT and the ST directions from the L-LT and the L-ST planes. The grains that were < 10 μm in length in the L direction were excluded from the analysis. The grains that do not contain internal substructure were identified as recrystallised grains for the measurement of area fraction. In this case, the high angle grain boundaries were firstly identified based on a boundary misorientation of > 15°, then the sub-grain structure classified according to a misorientation of < 2° using the HKL Channel 5 software. Further, the length densities and relative fractions of high angle (> 15°) and low angle (2° – 15°) grain boundaries were measured from EBSD grain boundary maps using a grey scale thresholding method utilising the Fiji software version 1.52p. The measurement was conducted in an area measuring 4 × 4 mm on the LT-ST plane for the specimens at the T/4 and the T/2 positions in T3 and T8 tempers. Orientation distribution functions (ODFs) were further calculated using the EBSD orientation datasets collected from the LT-ST plane using a harmonic series expansion with a truncation at $l_{max}$ = 22 from EBSD datasets. ODFs are present as plots of constant sections at $\phi_2$ = 45°, 65° and 90° in the Euler space defined by the Euler angles $\phi_1$, $\Phi$ and $\phi_2$ for a comparison between the T/4 and the T/2 positions in T3 and T8 alloys.

The fine precipitate and dispersoid phases were analysed using BSE imaging to characterise phase distribution at the T/4 and the T/2 positions in T3 and T8 tempers. Prior to the analysis, the polished specimens were further prepared using a Leica EM RES102 ion beam milling system with an $Ar^+$ source operated at 3 kV and 1.5 mA for 30 minutes to eliminate slight etching introduced by the chemo-mechanical polish. A FEI Magellan 400 XHR SEM equipped with a concentric backscatter (CBS) detector was then used at an accelerating voltage of +3 kV with a stage bias of –2 kV to acquire images with a pixel size of 7.3 nm over areas measuring 15.0 × 13.8 μm. The maximum depth of BSE was also estimated to be 11.3 nm for a 1 kV landing energy of electron on an Al substrate using the CASINO simulation software version 2.51.

To further characterise the phase distribution through the thickness of T8 alloy, twenty-five BSE images were collected from the regions measuring 2 × 2 mm at the T/4 and T/2 positions using the identical acquisition condition from the same locations as where EBSD analysis was performed. These images were segmented to identify the phases present using a feature-based classification method utilising the Trainable Weka Segmentation plug-in [24] in Fiji. This method classifies greyscale images into the sub-classes of pixels sharing similar visual characteristics using machine learning algorithms. The procedure for the use of Trainable Weka Segmentation has been previously detailed elsewhere as in [24,25]. In this case, the algorithm for image segmentation used the training features including Gaussian blur, Sobel filter, Hessian, membrane projections, difference of gaussians, variance, bilateral, Kuwahara and neighbours. The phases of interest were then classified using the FastRandomForest algorithm for the measurement of population density and average size using the Fiji software. The particles < 100 nm or with an aspect ratio < 2 were excluded from the analysis to highlight the needle-shaped $T_1$ phases distributed on grain boundaries. The volume fraction of $T_1$ phases was then calculated using the equation [26]:

$$f_v = \frac{N\pi t D^2}{4} \tag{4}$$

Where $f_v$ is the volume fraction, $N$ is the population density, $D$ is the average diameter and $t$ is the average thickness of $T_1$ phases.

Specimens for transmission electron microscopy (TEM) were extracted from the T/2 position of T3 and T8 alloys to compare between different tempers and the T/4 position of T8 alloy for a comparison of phase distribution through the plate thickness. The thin-foil specimens were prepared by grinding to < 100 μm in thickness, followed by twin jet electropolishing in a solution of 1:3 nitric acid to methanol at a temperature of –25 °C and a potential of 20 V. The precipitate phases were studied using a Thermo Scientific Talos F200X TEM at 200 kV. Selected area electron diffraction patterns (SAEDPs) were first collected from the $<110>_{Al}$ zone axis for the materials at the T3 and T8 states. TEM micrographs were then collected in the dark field (DF) mode from the $<110>_{Al}$ zone axis to reveal the fine precipitate and dispersoid phases using the superlattice reflections corresponding to the δ′ ($Al_3Li$)/θ′ ($Al_2Cu$)/β′ ($Al_3Zr$) and the edge-on $T_1$ phases. The misorientation of grain boundaries was also measured using SAEDPs with the procedure as detailed in [27,28].

In addition, the thickness of thin-foil specimens from the T8 alloy was measured using convergent beam electron diffraction (CBED) with the procedure as detailed in [29].

CBED patterns were collected from eight random locations to obtain average values of foil thickness for the specimens at the T/4 and T/2 positions. In this case, the specimens from the T/4 and the T/2 positions measure 48.8 ± 2.0 nm and 60.1 ± 6.6 nm, respectively. To investigate the population density and average size of precipitate phases, DF-TEM micrographs were further collected with a pixel size of ~ 0.4 nm from five regions with each covering an area of 1600 × 1600 nm for a statistical analysis of precipitates within the grain interiors. The population density and average size of precipitates were measured after segmentation of the particles using a grey scale thresholding method utilising the ImageJ software. The smallest δ′/θ′ and $T_1$ phases that were included for the analysis measure 3.1 nm in diameter and 2.6 nm in length, respectively. The volume fraction of $T_1$ precipitates were further calculated using the equation (4).

## 3. Results

### 3.1 Anisotropy of tensile strength

The mechanical behaviour of the Al-Li plate was evaluated using uniaxial tensile testing. Table 2 displays the tensile characteristics with the representative stress-strain curves as measured with the stress loading direction parallel with the L and LT directions shown in Figure 2.

**Table 2. Tensile strengths and elongation associated with the uniaxial tensile curves measured for the T3 and T8 specimens at the T/4 and T/2 positions. The global 0.2% offset strength is used as the yield strength.**

| Specimen | | T/4-T3 | | T/4-T8 | | T/2-T3 | | | T/2-T8 | | |
|---|---|---|---|---|---|---|---|---|---|---|---|
| Principal stress direction | | L | LT | L | LT | L | LT | ST | L | LT | ST |
| $\sigma_{p0.2}$ (MPa) | test 1 | 345 | 315 | 434 | 410 | 385 | 332 | 283 | 505 | 446 | 399 |
| | test 2 | 337 | 307 | 444 | 400 | 381 | 332 | 274 | 506 | 449 | 409 |
| | **average** | **341** | **311** | **439** | **405** | **383** | **332** | **279** | **506** | **448** | **404** |
| $\sigma_{UTS}$ (MPa) | test 1 | 414 | 417 | 495 | 500 | 461 | 433 | 408 | 560 | 536 | 497 |
| | test 2 | 420 | 417 | 495 | 500 | 455 | 433 | 412 | 571 | 538 | 506 |
| | **average** | **417** | **417** | **495** | **500** | **458** | **433** | **410** | **566** | **537** | **502** |
| $\varepsilon_f$ (%) | test 1 | 21.1 | 18.3 | 19.6 | 16.4 | 11.5 | 14.8 | 10.1 | 10.7 | 10.4 | 6.5 |
| | test 2 | 17.7 | 20.3 | 11.0 | 10.0 | 10.3 | 17.2 | 13.6 | 10.6 | 12.0 | 5.0 |
| | **average** | **19.4** | **19.3** | **15.3** | **13.2** | **10.9** | **16.0** | **11.9** | **10.7** | **11.2** | **5.8** |

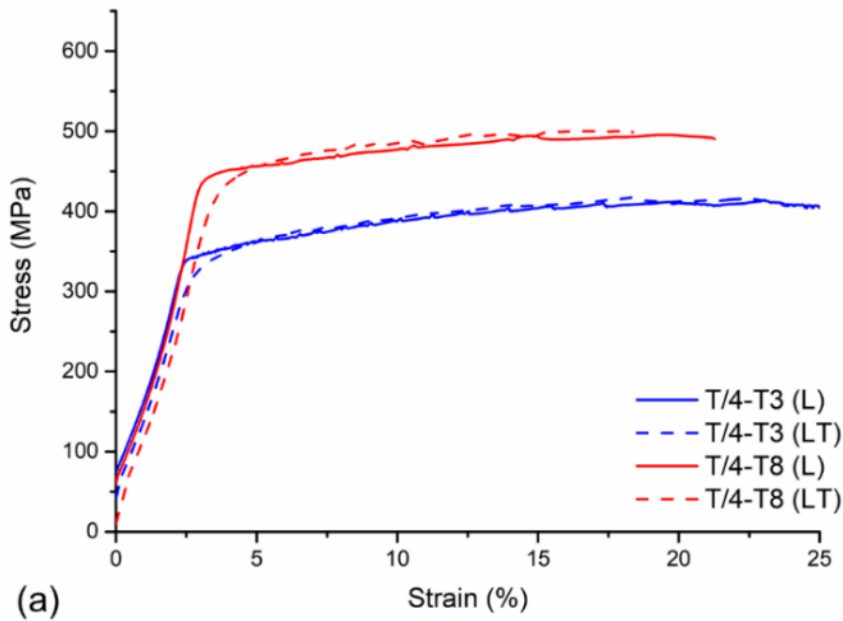

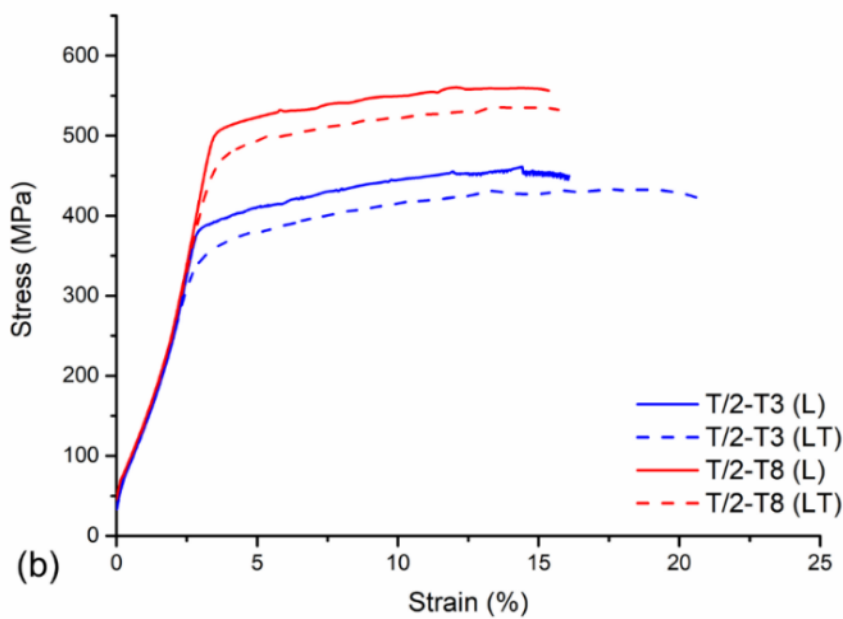

**Figure 2. A graph comparing the examples of stress-strain curves for (a) T3 and (b) T8 specimens at the T/4 and T/2 positions with stress loading direction parallel with the L and LT directions.**

Tensile testing shows that the tensile strength of T8 specimens is considerably higher than T3 specimens, whilst the elongation of T8 specimens is generally lower. It is also revealed that tensile strength is higher at the T/2 position than the T/4 position for both tempers. The tensile curves further highlight the difference in strength between the L and LT directions at the T/2 position, whilst the strengths are not significantly varied between the L and LT directions at the T/4 position.

The in-plane and through-thickness anisotropy of tensile strengths were further evaluated based on the measured values of tensile strengths using equations (2) and (3). Two tests were conducted to obtain average value and standard deviation for a comparison between T3 and T8 tempers, Figure 3.

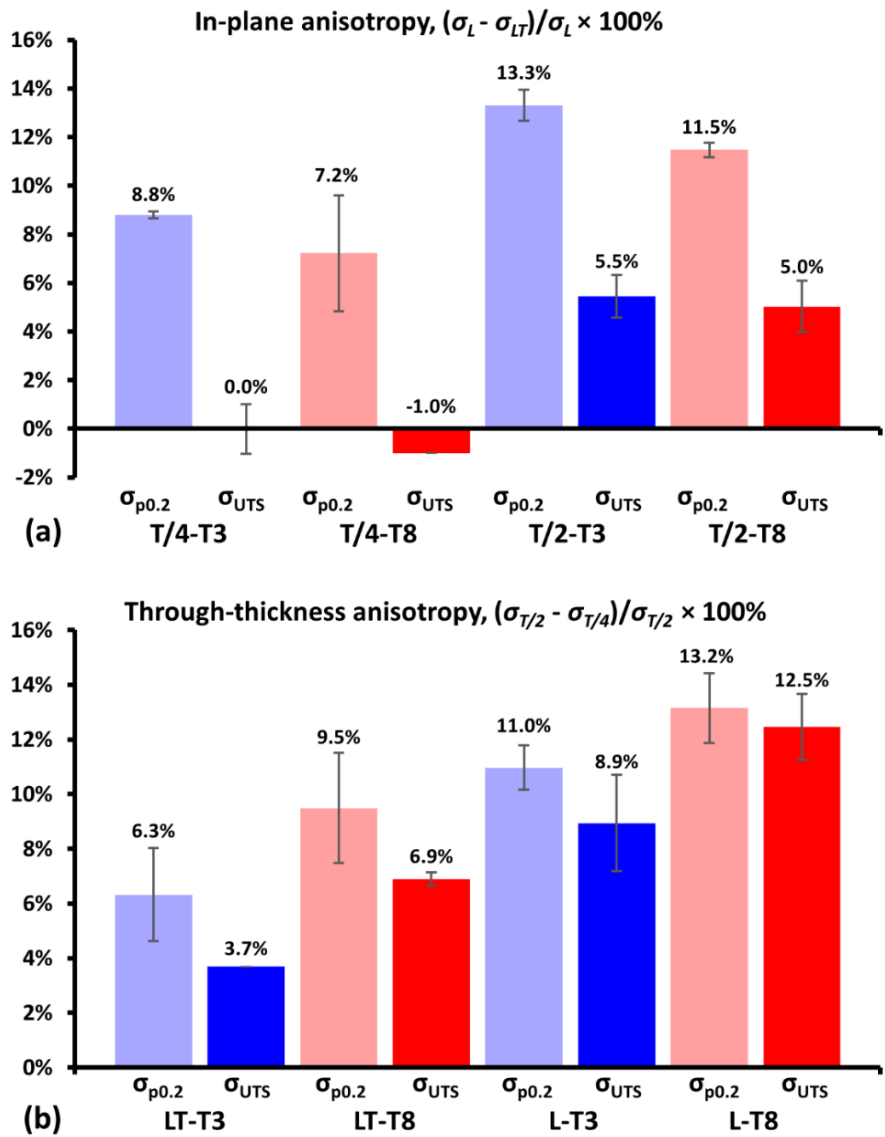

**Figure 3. Graphs comparing the average values of (a) in-plane and (b) through-thickness anisotropy between T3 and T8 tempers. The anisotropy was calculated based on the percentage of variation for tensile strengths between the L and LT directions (in-plane anisotropy) or the T/4 and T/2 positions (through thickness anisotropy).**

Figure 3a shows a higher level of in-plane anisotropy of tensile strengths at the T/2 position than the T/4 position in both T3 and T8 tempers. For instance, the in-plane anisotropy calculated based on the yield and ultimate strengths measure 8.8 ± 0.1% and 0.0 ± 1.0% at the T/4 position of T3 temper, whilst they measure 13.3 ± 0.6% and 5.5 ± 0.9% at the T/2 position. In addition, the in-plane anisotropy is slightly higher for the T3 temper than T8 at both the T/4 and T/2 positions. This is evident from comparing the anisotropy measuring 13.3 ± 0.6% and 5.5 ± 0.9% at the T/2 position of T3 temper with 11.5 ± 0.3% and 5.0 ± 0.1% at the same thickness position of T8 temper. Further, Figure 3b shows a higher level of through-thickness anisotropy for the strengths measured from the L direction than the LT direction. Importantly, through-thickness anisotropy is considerably higher for T8 temper than T3 in both the L and LT directions. The through-thickness anisotropy measured in the L direction gives 11.0 ± 0.8%, 8.9 ± 1.8% and 13.2 ± 1.3%, 12.5 ± 1.2% for T3 and T8 tempers, respectively.

*3.2 EBSD orientation mapping and grain structure analysis*

Figure 4 shows a three-dimensional overview of grain structure as revealed by EBSD mapping for the T3 and T8 alloys at T/4 and T/2 positions. It is evident that the grain structures in T3 and T8 alloys are predominantly composed of pancake-shaped grains that are elongated along the L direction. No significant variation in grain morphology was observed between the different thickness positions in both tempers. The average sizes of grain measure ~ 150 μm, ~ 110 μm and ~ 10 μm in the L, LT and ST directions, respectively. No significant variation in grain dimensions was observed between the T/4 and the T/2 positions, indicating a reasonably homogeneous grain structure through the thickness of both tempers. The maximum dimensions of grain are > 2,000 μm in the L direction, > 1,000 μm in the LT direction and > 80 μm in the ST direction. The statistical distribution of grain dimension is further illustrated in detail in supporting information, Figure S1.

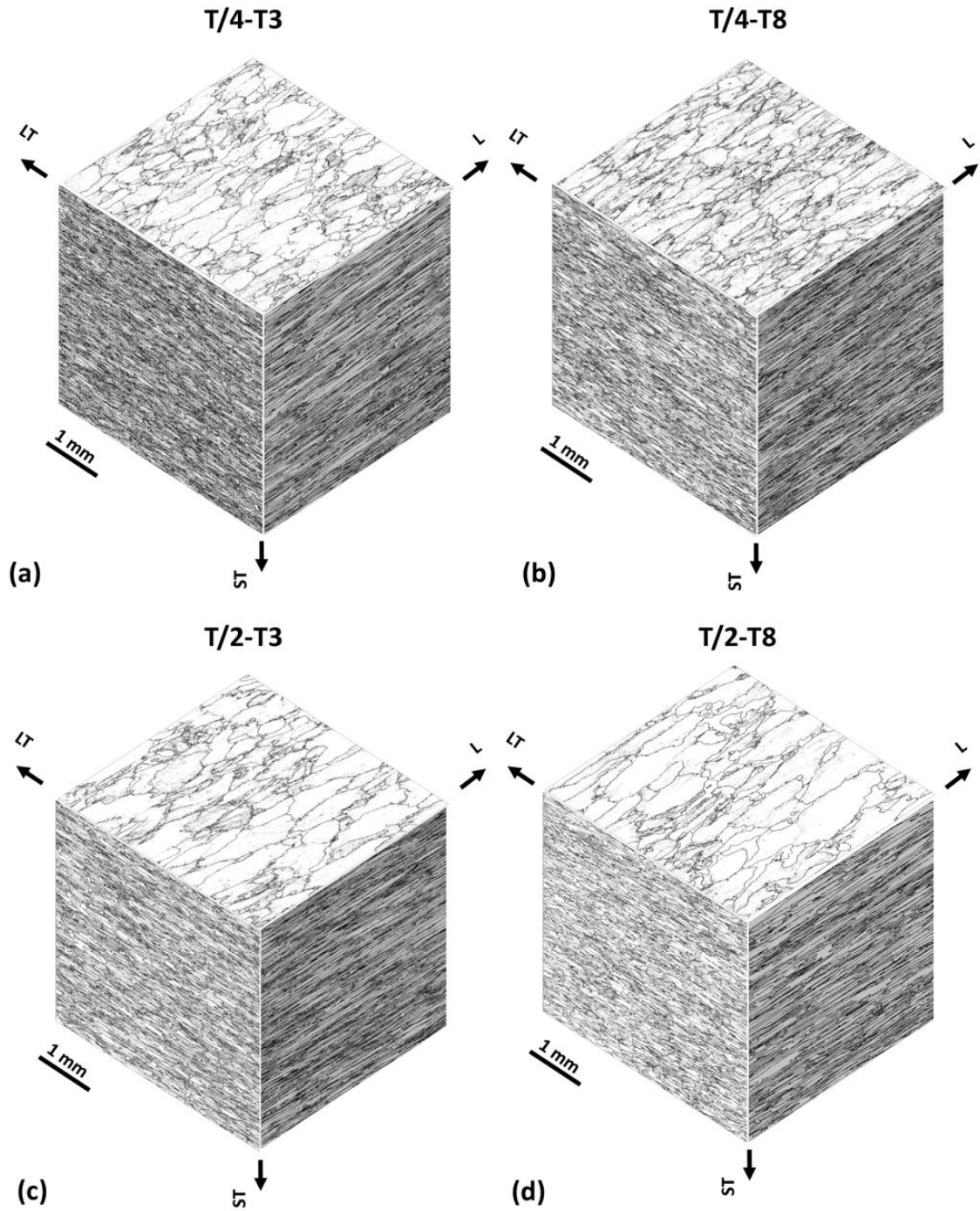

**Figure 4. EBSD grain boundary maps showing a three-dimensional overview of grain structure with grain boundaries (> 15°) outlined for specimens at (a, b) the T/4 and (c, d) the T/2 positions in (a, c) T3 and (b, d) T8 tempers.**

Figure 5 shows the EBSD grain boundary maps comparing grain structures on the LT-ST plane between the different thickness positions in T3 and T8 alloys. Table 3 further presents the quantitative characteristics of recrystallised grains and grain boundaries.

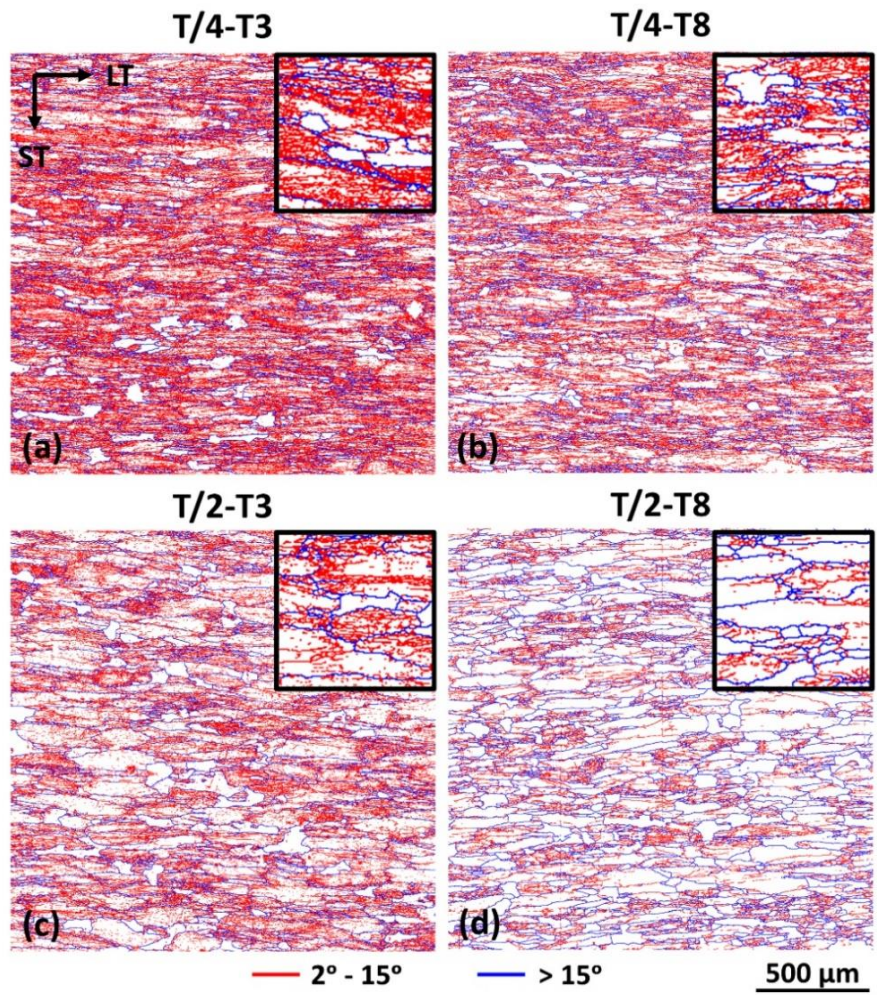

**Figure 5.** EBSD grain boundary maps showing the grain structure from the LT-ST plane at (a, b) the T/4 and (c, d) the T/2 positions in (a, c) T3 and (b, d) T8 tempers. The low angle (2° - 15°) and high angle (> 15°) grain boundaries were outlined by solid lines in red and blue, respectively. The insets (250 × 250 μm) show the details of grain structure in a magnified view.

**Table 3.** The area fraction of recrystallised grain, the lengths per unit area and relative fractions of low angle (2° - 15°) and high angle (> 15°) grain boundaries at the T/4 and the T/2 positions in T3 and T8 tempers.

| Specimens | Area fraction of recovered or recrystallised grains (%) | Low angle grain boundary (2° - 15°) | | High angle grain boundary (> 15°) | |
|---|---|---|---|---|---|
| | | Boundary length (mm/mm$^2$) | Relative fraction | Boundary length (mm/mm$^2$) | Relative fraction |
| T/4-T3 | 3.9 | 209.3 | 0.80 | 53.4 | 0.20 |
| T/4-T8 | 6.7 | 132.8 | 0.73 | 49.1 | 0.27 |
| T/2-T3 | 5.5 | 116.9 | 0.74 | 40.4 | 0.26 |
| T/2-T8 | 10.0 | 48.5 | 0.55 | 39.6 | 0.45 |

Figure 5 reveals the substructure as identified by low angle grain boundary (i.e. 2° - 15° in misorientation) within the pancake-shaped grains and a small population of grains containing no internal substructure. The presence of low angle grain boundary is due to the accumulation of dislocations within the grain interiors after thermomechanical processing and is influenced by the deformation and degree of recovery [30,31]. The quantitative evaluation further shows that the area fraction of the grains without internal substructure (determined to be recovered or recrystallised grains) is slightly lower at the T/4 position than the T/2 position and is higher in T8 temper than T3 temper. In addition, the statistics of grain boundary characteristics show that the length densities and relative fractions of high angle grain boundary (i.e. > 15° in misorientation) are not significantly varied through the plate thickness in both tempers. However, the length densities and relative fractions of low angle grain boundary are evidently lower at the T/2 position than the T/4 position, and lower in T8 temper than T3 temper.

Figure 6 displays the crystallographic texture at the T/4 and the T/2 positions in the T3 and T8 alloys by using the ODF sections with constant values of the Euler angle $\phi_2 =$ 45°, 60° and 90°, revealing the rolling texture components distributed on the β-fibre orientations. The ODF sections calculated from the T/4 position, Figures 7a and 7b, display a low intensity of texture components, with the maximum orientation density (i.e. $f \approx 6$) close to the {4 4 11}<11 11 8> D orientations. Figures 7c and 7d reveal that the β-fibre textures are more developed at the T/2 position. Strong β-fibre texture was observed with the maximum intensity (i.e. $f \approx 30$) close to the {011}<211> Bs orientation, followed by the {123}<634> S orientation in both T3 and T8 tempers. The statistical analysis of crystal orientation does not reveal significant crystallographic texture contributions from recrystallisation (e.g. $f <$ 3 at the {001}<100> Cube orientation) across all specimens. In addition, no significant variation of crystallographic texture was observed between T3 and T8 tempers at the same thickness positions.

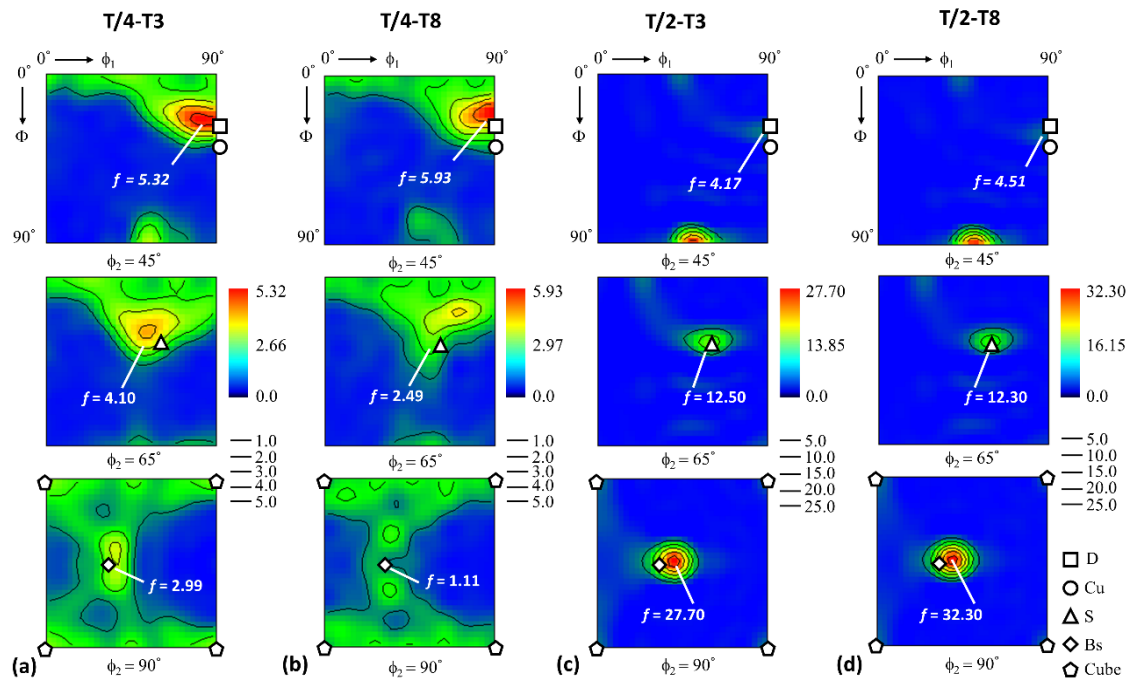

**Figure 6.** Sections through ODFs at the Euler angle $\phi_2 = 45°$, 65° and 90° at (a, b) the T/4 and (c, d) the T/2 positions in (a, c) T3 and (b, d) T8 tempers. Symbols indicate the locations of ideal orientation for the relevant texture components comprising the β fibre. Numbers indicate the ODF intensities calculated at the orientations as indicated by white lines.

*3.3 Precipitate and dispersoid phases*

Figure 7 shows the BSE micrographs comparing the distribution of precipitate and dispersoid phases between T3 and T8 tempers at the T/4 and the T/2 positions observed on the LT-ST plane.

Figures 7a and 7d reveal the coarse needle-shaped $T_1$ and the round-shaped dispersoid phases at the T/4 and the T/2 positions in T3 alloy. The coarse $T_1$ precipitates measuring 3 – 5 μm in length and ~ 100 nm in width are preferentially distributed on high angle grain boundaries. The round-shaped dispersoid particles measuring 100 – 300 nm in diameter are predominantly the $Al_{20}Cu_2Mn_3$ phases as identified using EDX. These particles are distributed either on grain boundaries or within the grain interiors. In addition, the dispersoid particles measuring < 50 nm in diameter were observed within the grain interiors. These particles are further identified to be the β′ ($Al_3Zr$) dispersoid phases using DF-TEM as illustrated in the following sections.

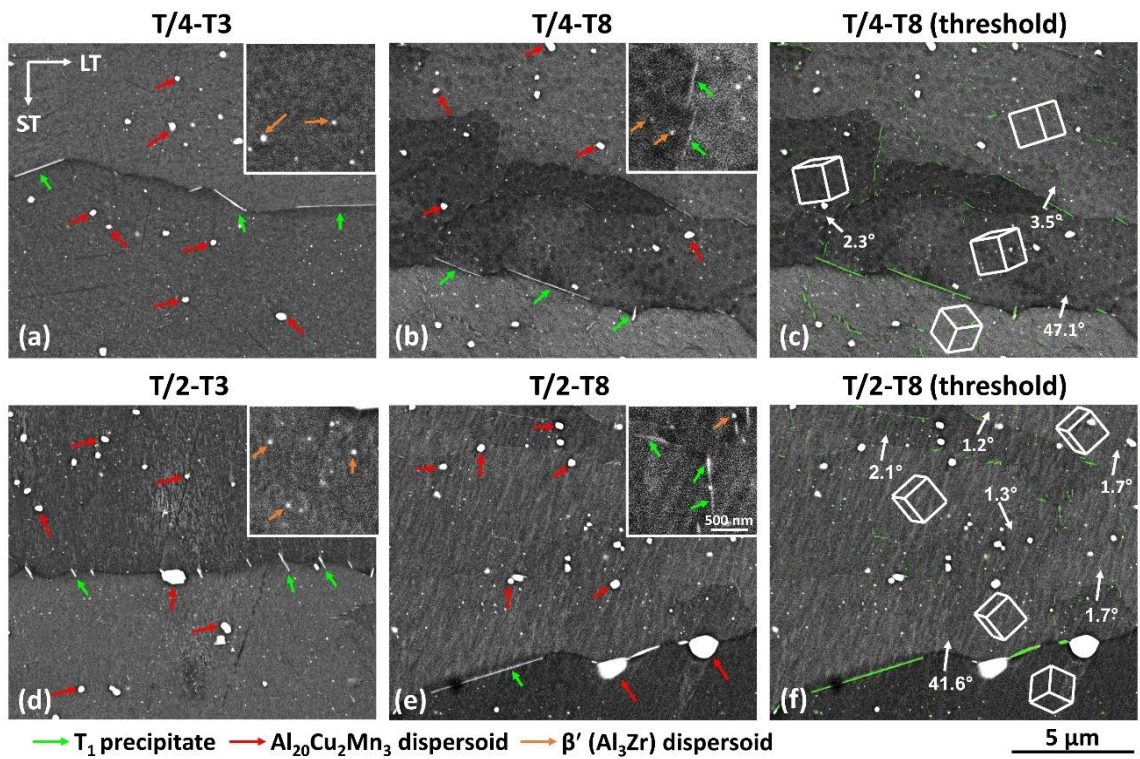

**Figure 7.** BSE micrographs showing the distribution of precipitate and dispersoid phases at (a – c) the T/4 and (d – f) the T/2 positions in the (a, d) T3 and (b, c, e, f) T8 alloys. The insets (1.8 × 1.8 μm) in (b, e) exhibit the details of needle-shaped $T_1$ precipitate and round-shaped dispersoid particles in a magnified view. The phases are indicated by arrows: $T_1$ precipitate (green), the $Al_{20}Cu_2Mn_3$ (red) and the β′ ($Al_3Zr$) dispersoids (orange). The $T_1$ particles in (b) and (e) are further labelled in green using the Trainable Weka Segmentation plug-in as illustrated in (c) and (f). The overlays and arrows in (c) and (f) indicate the crystal orientation and misorientation of grain boundaries measured by EBSD, respectively.

Figures 7b and 7e reveal similar phase constituents in T8 alloy at the T/4 and the T/2 positions. In addition, fine particles of $T_1$ precipitates were observed on the low angle grain boundaries of internal substructure within the grain interiors. These $T_1$ precipitate as observed in T8 temper only are formed during the artificial ageing of T8 treatment, whilst the coarse $T_1$ precipitates observed in both tempers are formed during quenching upon a slow cooling rate at the plate centre [20].

The coarse quench-induced and the ageing-induced $T_1$ precipitates on grain boundary were highlighted using the Trainable Weka Segmentation plug-in for a comparison of phase distribution between the T/4 and the T/2 positions in T8 temper, as shown in Figures 7c and 7f. The intergranular $T_1$ precipitates are not significantly varied in size between the T/4 and T/2 positions. However, the population density of $T_1$ precipitates measures ~ $7.6 \times 10^{13}$ /mm$^3$ and ~ $5.4 \times 10^{13}$ /mm$^3$ at the T/4 and T/2 positions, indicating considerable variation through the plate thickness.

The fine dispersoids and ageing-induced precipitates were further characterised using TEM. SAEDPs and DF-TEM micrographs were first collected from the T/2 position for a comparison between T3 and T8 tempers, Figure 8.

The SAEDP collected from T3 temper shows diffuse spots for the superlattice reflections corresponding to δ′($Al_3Li$) /θ′($Al_2Cu$) precipitates and β′ dispersoids [4,32], Figure 8a. The superlattice reflections for δ′/θ′/β′ phases are more intense in the SAEDP collected from T8 temper, Figure 8b. Figure 8b also shows the characteristic spots and streaks that correspond to the superlattice reflection of $T_1$ precipitates [4,32]. Figure 8c shows the δ′/θ′ precipitates measuring < 5 nm in diameter within the grain and sub-grain interiors of T3 specimen. Round β′ dispersoids were also observed measuring < 50 nm in diameter. No precipitates were observed at low angle grain boundaries in T3 specimen, which is consistent with the observations from BSE images shown in Figure 7.

Figure 8d further reveals the δ′/θ′ precipitates measuring 5 – 30 nm within the grain and sub-grain interiors of T8 specimens, whilst the depletion of precipitates was observed in the vicinities of grain boundary. The precipitate free zones (PFZs) in T8 alloy are typically ~ 100 nm and ~ 30 nm in width along high angle and low angle boundaries, respectively. Figure 8e further reveals the needle-shaped $T_1$ precipitates within the grain and sub-grain interiors and at low angle grain boundaries in T8 specimen. The intragranular $T_1$ precipitates are < 50 nm in length, whilst the intergranular $T_1$ precipitates are typically 100 – 200 nm in length. This is consistent with the observations from BSE imaging as shown in Figure 7. The intergranular $T_1$ precipitates are relatively larger in size due to an enhanced diffusion rate arising from the grain boundary pipe diffusion [33].

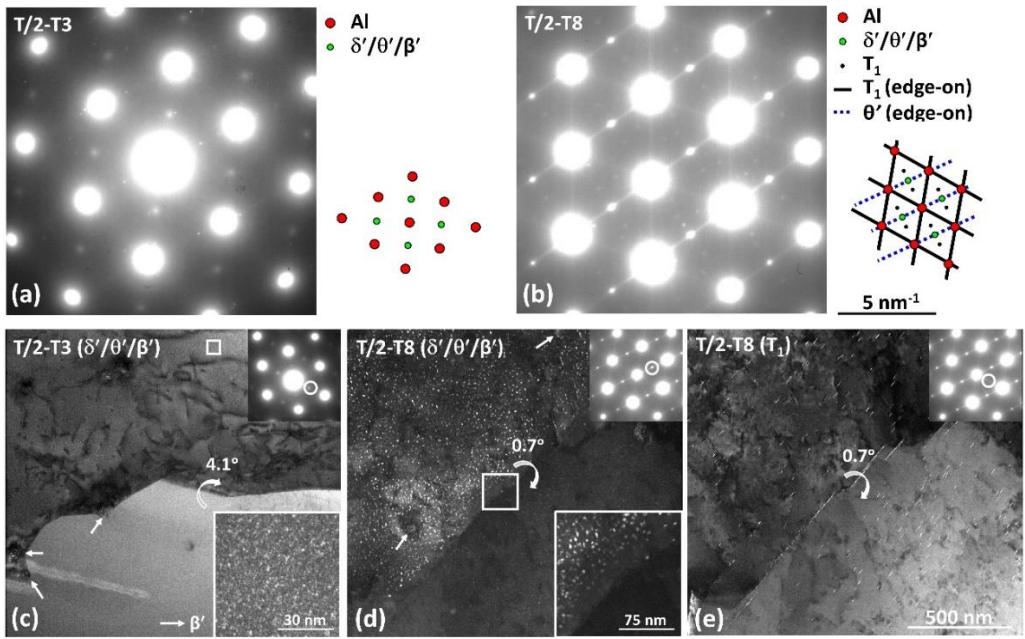

**Figure 8.** SAEDPs collected from the <110>$_{Al}$ zone axis at the T/2 position of (a) T3 and (b) T8 alloys with the illustrations of superlattice reflection from precipitate and dispersoid phases. DF-TEM micrographs from (c) T3 and (d, e) T8 tempers are also included to demonstrate precipitate and dispersoid phases using different superlattice reflections. The insets in (c – e) show the reflections used for DF-TEM imaging as circled in SAEDPs and the details of precipitates within the grain interior and at grain boundary. The β′ dispersoids are indicated by white arrows in (c, d).

The intragranular precipitate and dispersoid phases within the grain and sub-grain interiors were further characterised in the DF-TEM mode to investigate phase distribution through the plate thickness of T8 alloy, Figure 9.

Figure 9a reveals the fine δ′/θ′ precipitate and the round-shaped β′ dispersoid phases at the T/4 position of T8 alloy. The δ′/θ′ precipitates are either spherical or lenticular in shape, with the longitudinal axis of the lenticular particles parallel to one of the {100}$_{Al}$ plane variants. This is consistent with the observations of δ′/θ′ precipitates in a 2099-T8 alloy as previously reported in [5]. The spherical-shaped phases have been confirmed to be the δ′ precipitates, whilst the lenticular particles are a combination of δ′ formed on the edge-on θ′ phases in parallel with the {100}$_{Al}$ plane variants [5]. Figure 9b further reveals the needle-shaped T$_1$ precipitates using the corresponding superlattice reflection.

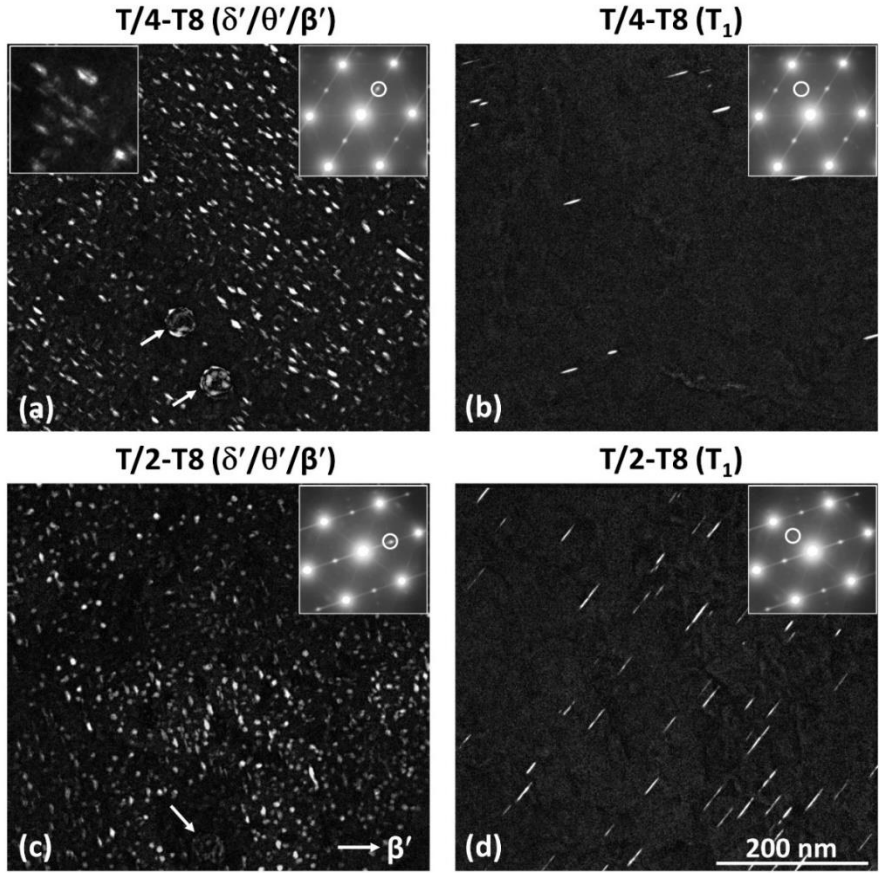

**Figure 9. DF-TEM micrographs collected with the superlattice reflections of (a, c) the δ′/θ′/β′ precipitates and (b, d) the edge-on $T_1$ precipitates at (a, b) the T/4 and (c, d) the T/2 positions of T8 alloy. The insets show the reflections used for DF-TEM imaging as circled in SAEDPs and the precipitates in a magnified view measuring 60 × 60 nm. The white arrows indicate the β′ dispersoids.**

Figures 9c and 9d exhibit the identical phase constituents at the T/2 position of T8 alloy. The population density and average size of δ′/θ′ precipitates are not significantly varied between the T/4 and the T/2 positions. However, the population density of intragranular $T_1$ precipitates measures $1.9 \times 10^{21}$ and $3.7 \times 10^{21}$ /mm$^3$ at the T/4 and the T/2 positions, indicating significant variation through the plate thickness. The volume fraction of intragranular $T_1$ precipitates was further calculated to be $2.0 \times 10^{-3}$ and $2.8 \times 10^{-3}$ at the T/4 and the T/2 positions, respectively.

## 4. Discussion

*4.1 Through-thickness variation of microstructure and crystallographic texture*

A detailed metallographic examination of through-thickness microstructure for this 50-mm thick plate Al-Li alloy in two different tempers has been conducted. In this case, the pancake-shaped morphology of grain structure is as expected for the Al-Li rolled plates containing an unrecrystallised microstructure [3,17,18]. The EBSD results shown in Figure 4 further confirm the similarity of grain size between the T/4 and T/2 positions in both tempers, indicating a reasonably homogeneous grain structure through the plate thickness. However, the T/2 position shows a coarser substructure and more pronounced crystallographic texture as compared to the T/4 position. The distribution of $T_1$ precipitates is also varied through the plate thickness.

The through-thickness gradient of temperature and deformation strain during hot rolling causes the variation of substructure and crystallographic texture. As previously reported [34–36], the plate surface experiences a faster cooling and higher strain than the plate centre due to heat transfer and the presence of shear strain. This usually leads to a more refined grain structure and/or substructure (e.g. dislocation cells or sub-grains) within the grain interiors [30,31], as observed at the T/4 position. The through-thickness gradient of crystallographic texture, as indicated in Figure 6, is consistent with observations from the existing studies conducted on thick-section Al-Cu-Li plates [9,17]. In this case, the strong β-fibre texture present at the plate centre is typical of the hot-rolled Al sheets and plates as a result of plane strain compression [39,40]. The maximum orientation density was observed at the Bs orientations due to the highest orientation stability at the corresponding deformation conditions at the plate centre [39]. The less pronounced crystallographic texture present at the T/4 position was caused by the deviation of deformation strain from plane strain compression as indicated in [16].

The variation of substructure as identified between T3 and T8 tempers is related to the different ageing treatments conducted. In this case, the artificial ageing conducted at 120 and 145 °C induced the static recovery of the cold-worked structure formed after pre-stretch in T8 alloy. This is consistent with observations from the existing studies showing static recovery of Al alloys after artificial ageing [37,38]. The through-thickness variation of substructure further leads to an inhomogeneous distribution of ageing-induced $T_1$ precipitates as indicated in Figures 7 and 9. In this case, the population density of the ageing-induced intergranular $T_1$ precipitates is ~ 30% higher at the T/4 position than the T/2 position in T8 temper. This is in a good agreement with the relative fraction of low angle grain boundaries that is ~ 30% higher at the T/4 position. $T_1$ precipitates are known to nucleate preferentially on the sites such as the interface of pre-existing secondary phase particles, dislocations and grain boundaries [47,48]. The higher length density and relative fraction of low angle grain boundary provides more sites available for the nucleation of $T_1$ precipitates during artificial ageing and, in turn, leads to a higher population of intergranular $T_1$ precipitates at the T/4 position. This consequently contributes to a lower population density of intragranular $T_1$ precipitates at the T/4 position.

*4.2 Influence of microstructure and crystallographic texture on mechanical anisotropy*

The alloys being investigated in this study generally have comparable tensile strengths and elongation as compared to the existing Al-Li-Cu-Mg plate materials [1,49,50]. The T8 alloy shows higher strength than T3 alloy due to a large amount of strengthening precipitates (e.g. the $\delta'$ and $T_1$ phases) that were formed during artificial ageing [4]. The presence of PFZs along grain boundary further contributes to the lower elongation of T8 alloy by introducing strain localisation in adjacent to the unshearable intergranular $T_1$ precipitates [51].

The similarity of grain size and morphology between the T/4 and the T/2 positions suggests a minor influence on the through-thickness anisotropy of tensile strengths. However, the pancake-shaped grain morphology contributes to the in-plane anisotropy due to the variation of effective slip length for the <110>{111} slip systems between the L and the LT directions of pancake-shaped grains [52]. In addition, the tensile strengths and elongation are the lowest at the T/2 position in the ST direction for both tempers. This reflects the small grain dimension in this direction and, hence, a higher length density of grain boundaries and the adjacent PFZs being loaded perpendicularly.

The strong crystallographic texture is the main cause of in-plane mechanical anisotropy at the T/2 position. As previously reported in [9,41,53], the presence of texture components at the {011}<211> Bs orientation is commonly associated with strong in-plane anisotropy with higher strengths along the L compared to the LT direction for Al-Li-Cu-Mg alloys. The theoretical calculation of Schmid and Taylor factors further suggests the lowest strength at ~ 50 – 60° to the L direction [53,54]. In comparison to the T/2 position, the difference of tensile strengths was reduced between the L and the LT directions at the T/4 position. This indicates a lower level of in-plane anisotropy resulted from the less pronounced crystallographic texture at the T/4 position.

The in-plane anisotropy of T8 alloy was slightly reduced as compared to T3 alloy. This may be caused by the static recovery of cold-worked structure during artificial ageing as indicated in Figure 5. Static recovery has been shown to cause the formation of sub-grains that are less elongated in shape and larger in size [31,55]. This is considered to mitigate the influence of pancake-shaped grain morphology and, in turn, reduces in-plane anisotropy. The formation of fine precipitates during artificial ageing is generally expected to increase in-plane anisotropy. For instance, the formation of coherent shearable δ′ precipitates increases in-plane anisotropy by promoting slip planarity [1]. The inhomogeneous deformation in strongly textured 2090 and 2198 alloys also causes an uneven distribution of $T_1$ precipitates on the $\{111\}_{Al}$ plane variants, which, in turn, increases in-plane anisotropy [11,12]. However, in this case, the increase of in-plane anisotropy was not observed in T8 temper after artificial aging. This suggests a homogeneous distribution of $T_1$ precipitates on the $\{111\}_{Al}$ plane variants as a result of a sufficient extent of cold working (i.e. ~ 4%) as introduced during pre-stretch operation [52].

The inhomogeneous distribution of $T_1$ precipitates further contributes to the through-thickness anisotropy of tensile strengths in T8 alloy. These precipitates deliver the strengthening effect via a shearing mechanism unless substantially coarsened to large, unshearable particles leading to Orowan by-passing of dislocations [56]. Based on the quantitative evaluation of $T_1$ precipitate distribution as illustrated in Figures 7 and 9, the volume fraction of the ageing-induced intragranular $T_1$ precipitates is ~ 40% higher at the T/2 position than the T/4 position. This indicates a significantly higher strength contribution from these precipitates at the T/2 position [58], leading to a higher level of through-thickness anisotropy of T8 alloy.

**5. Conclusions**

- The 50-mm thick-section Al-Li alloys containing an unrecrystallised matrix with

pancake-shaped grain morphology exhibit higher tensile strengths for the T8 temper than the T3 temper due to a higher level of ageing-induced precipitates including $\delta'$, $\theta'$ and $T_1$ phases.

- Both T3 and T8 tempers exhibit considerable in-plane anisotropy at the plate centre due to strong crystallographic texture with the β-fibre orientation. In-plane anisotropy is ~ 5% lower at the T/4 position with less-pronounced crystallographic texture.
- Through-thickness anisotropy is ~ 3% higher in T8 temper than T3 temper. This is due to the ageing-induced intragranular $T_1$ precipitates that are ~ 40% higher in volume fraction at the plate centre.
- The variation of ageing-induced $T_1$ precipitates is related with the through-thickness inhomogeneity of substructure arising from the strain gradient during hot rolling and static recovery of substructure during artificial aging.

**Acknowledgements**

This work was supported by the Henry Royce Institute for Advanced Materials, funded through EPSRC grants EP/R00661X/1.

**References**

[1]   N.E. Prasad, A.A. Gokhale, R.J.H. Wanhill, Aluminium–lithium alloys, in: Aerosp. Mater. Mater. Technol., Springer, 2017: pp. 53–72.

[2]   J.C. Williams, E.A. Starke Jr, Progress in structural materials for aerospace systems, Acta Mater. 51 (2003) 5775–5799.

[3]   R.J. Rioja, Fabrication methods to manufacture isotropic Al-Li alloys and products for space and aerospace applications, Mater. Sci. Eng. A. 257 (1998) 100–107.

[4]   S.C. Wang, M.J. Starink, Precipitates and intermetallic phases in precipitation hardening Al–Cu–Mg–(Li) based alloys, Int. Mater. Rev. 50 (2005) 193–215.

[5]   Y. Ma, X. Zhou, G.E. Thompson, T. Hashimoto, P. Thomson, M. Fowles, Distribution of intermetallics in an AA 2099-T8 aluminium alloy extrusion,

Mater. Chem. Phys. 126 (2011) 46–53.

[6] S.J. Andersen, C.D. Marioara, J. Friis, S. Wenner, R. Holmestad, Precipitates in aluminium alloys, Adv. Phys. X. 3 (2018) 1479984.

[7] K. V Jata, A.K. Hopkins, R.J. Rioja, The anisotropy and texture of Al-Li alloys, in: Mater. Sci. Forum, Trans Tech Publ, 1996: pp. 647–652.

[8] A.A. El-Aty, Y. Xu, X. Guo, S.-H. Zhang, Y. Ma, D. Chen, Strengthening mechanisms, deformation behavior, and anisotropic mechanical properties of Al-Li alloys: A review, J. Adv. Res. 10 (2018) 49–67.

[9] S.J. Hales, R.A. Hafley, Texture and anisotropy in Al-Li alloy 2195 plate and near-net-shape extrusions, Mater. Sci. Eng. A. 257 (1998) 153–164.

[10] N.E. Prasad, S. V Kamat, K.S. Prasad, G. Malakondaiah, V. V Kutumbarao, In-plane anisotropy in the fracture toughness of an Al-Li 8090 alloy plate, Eng. Fract. Mech. 46 (1993) 209–223.

[11] N.J. Kim, E.W. Lee, Effect of T1 precipitate on the anisotropy of Al Li alloy 2090, Acta Metall. Mater. 41 (1993) 941–948.

[12] Z. Tian-Zhang, J. Long, X. Yong, Z. Shi-Hong, Anisotropic yielding stress of 2198 Al–Li alloy sheet and mechanisms, Mater. Sci. Eng. A. 771 (2020) 138572.

[13] A.A. Csontos, E.A. Starke, The effect of processing and microstructure development on the slip and fracture behavior of the 2.1 wt pct Li AF/C-489 and 1.8 wt pct Li AF/C-458 Al-Li-Cu-X alloys, Metall. Mater. Trans. A. 31 (2000) 1965–1976.

[14] R.J. Rioja, J. Liu, The evolution of Al-Li base products for aerospace and space applications, Metall. Mater. Trans. A. 43 (2012) 3325–3337.

[15] Y. Zuo, F.U. Xing, J. Cui, X. Tang, M.A.O. Lu, L.I. Lei, Q. Zhu, Shear deformation and plate shape control of hot-rolled aluminium alloy thick plate prepared by asymmetric rolling process, Trans. Nonferrous Met. Soc. China. 24 (2014) 2220–2225.

[16] S. Li, N. Qin, J. Liu, X. Zhang, Microstructure, texture and mechanical properties of AA1060 aluminum plate processed by snake rolling, Mater. Des. 90 (2016) 1010–1017.

[17] P. Wu, Y. Deng, J. Zhang, S. Fan, X. Zhang, The effect of inhomogeneous microstructures on strength and fatigue properties of an Al-Cu-Li thick plate, Mater. Sci. Eng. A. 731 (2018) 1–11.

[18] Z. Kuo, J. Liu, Y.U. Mei, S. Li, Through-thickness inhomogeneity of precipitate distribution and pitting corrosion behavior of Al–Li alloy thick plate, Trans. Nonferrous Met. Soc. China. 29 (2019) 1793–1802.

[19] K.E. Crosby, R.A. Mirshams, S.S. Pang, Development of texture and texture gradient in Al-Cu-Li (2195) thick plate, J. Mater. Sci. 35 (2000) 3189–3195.

[20] B.M. Gable, A.A. Csontos, E.A. Starke Jr, A quench sensitivity study on the novel Al–Li–Cu–X alloy AF/C 458, J. Light Met. 2 (2002) 65–75.

[21] Metallic Materials - Tensile Testing - Part 1: Method of Test at Room Temperature, Gen. Adm. Qual. Supervision, Insp. Quar. People's Repub. China Stand. Adm. People's Repub. China. (2010).

[22] R.K. Singh, A.K. Singh, N.E. Prasad, Texture and mechanical property anisotropy in an Al–Mg–Si–Cu alloy, Mater. Sci. Eng. A. 277 (2000) 114–122.

[23] H. Hu, X. Wang, Effect of heat treatment on the in-plane anisotropy of as-rolled 7050 aluminum alloy, Metals (Basel). 6 (2016) 79.

[24] I. Arganda-Carreras, V. Kaynig, C. Rueden, K.W. Eliceiri, J. Schindelin, A. Cardona, H. Sebastian Seung, Trainable Weka Segmentation: a machine learning tool for microscopy pixel classification, Bioinformatics. 33 (2017) 2424–2426.

[25] C. Lormand, G.F. Zellmer, K. Németh, G. Kilgour, S. Mead, A.S. Palmer, N. Sakamoto, H. Yurimoto, A. Moebis, Weka trainable segmentation plugin in imagej: A semi-automatic tool applied to crystal size distributions of microlites in volcanic rocks, Microsc. Microanal. 24 (2018) 667–675.

[26]   T. Dorin, A. Deschamps, F. De Geuser, C. Sigli, Quantification and modelling of the microstructure/strength relationship by tailoring the morphological parameters of the T1 phase in an Al–Cu–Li alloy, Acta Mater. 75 (2014) 134–146.

[27]   Q. Liu, A simple method for determining orientation and misorientation of the cubic crystal specimen, J. Appl. Crystallogr. 27 (1994) 755–761.

[28]   Q. Liu, A simple and rapid method for determining orientations and misorientations of crystalline specimens in TEM, Ultramicroscopy. 60 (1995) 81–89.

[29]   D.B. Williams, C.B. Carter, The transmission electron microscope, in: Transm. Electron Microsc., Springer, 1996: pp. 3–17.

[30]   M.T. Pérez-Prado, J.A. del Valle, J.M. Contreras, O.A. Ruano, Microstructural evolution during large strain hot rolling of an AM60 Mg alloy, Scr. Mater. 50 (2004) 661–665.

[31]   H.J. McQueen, E. Evangelista, Substructures in aluminium from dynamic and static recovery, Czechoslov. J. Phys. B. 38 (1988) 359–372.

[32]   K.S. Kumar, S.A. Brown, J.R. Pickens, Microstructural evolution during aging of an AlCuLiAgMgZr alloy, Acta Mater. 44 (1996) 1899–1915.

[33]   I. Kaur, W. Gust, Y. Mishin, Fundamentals of grain and interphase boundary diffusion, Wiley Chichester, 1995.

[34]   M.A. Wells, D.M. Maijer, S. Jupp, G. Lockhart, M.R. van der Winden, Mathematical model of deformation and microstructural evolution during hot rolling of aluminium alloy 5083, Mater. Sci. Technol. 19 (2003) 467–476.

[35]   H. Ahmed, M.A. Wells, D.M. Maijer, B.J. Howes, M.R. van der Winden, Modelling of microstructure evolution during hot rolling of AA5083 using an internal state variable approach integrated into an FE model, Mater. Sci. Eng. A. 390 (2005) 278–290.

[36] C. Ma, L. Hou, J. Zhang, L. Zhuang, Influence of thickness reduction per pass on strain, microstructures and mechanical properties of 7050 Al alloy sheet processed by asymmetric rolling, Mater. Sci. Eng. A. 650 (2016) 454–468.

[37] P. Ying, Z. Liu, S. Bai, J. Wang, J. Li, M. Liu, L. Xia, Effect of artificial aging on the Cu-Mg co-clustering and mechanical behavior in a pre-strained Al-Cu-Mg alloy, Mater. Sci. Eng. A. 707 (2017) 412–418.

[38] W.J. Poole, J.A. Sæter, S. Skjervold, A model for predicting the effect of deformation after solution treatment on the subsequent artificial aging behavior of AA7030 and AA7108 alloys, Metall. Mater. Trans. A. 31 (2000) 2327–2338.

[39] S. Li, F. Sun, H. Li, Observation and modeling of the through-thickness texture gradient in commercial-purity aluminum sheets processed by accumulative roll-bonding, Acta Mater. 58 (2010) 1317–1331.

[40] L.A.I. Kestens, H. Pirgazi, Texture formation in metal alloys with cubic crystal structures, (2016).

[41] S.J. Hales, Structure-Property Correlations in Al-Li Alloy Integrally Stiffened Extrusions, DIANE Publishing, 2001.

[42] A.K. Singh, G.G. Saha, A.A. Gokhale, R.K. Ray, Evolution of texture and microstructure in a thermomechanically processed Al-Li-Cu-Mg alloy, Metall. Mater. Trans. A. 29 (1998) 665–675.

[43] G. Itoh, Q. Cui, M. Kanno, Effects of a small addition of magnesium and silver on the precipitation of T1 phase in an Al 4% Cu 1.1% Li 0.2% Zr alloy, Mater. Sci. Eng. A. 211 (1996) 128–137.

[44] V. Araullo-Peters, B. Gault, F. De Geuser, A. Deschamps, J.M. Cairney, Microstructural evolution during ageing of Al–Cu–Li–x alloys, Acta Mater. 66 (2014) 199–208.

[45] D. Liu, Y. Ma, J. Li, R. Zhang, H. Iwaoka, S. Hirosawa, Precipitate microstructures, mechanical properties and corrosion resistance of Al-1.0 wt% Cu-2.5 wt% Li alloys with different micro-alloyed elements addition, Mater.

Charact. (2020) 110528.

[46] E. Gumbmann, F. De Geuser, C. Sigli, A. Deschamps, Influence of Mg, Ag and Zn minor solute additions on the precipitation kinetics and strengthening of an Al-Cu-Li alloy, Acta Mater. 133 (2017) 172–185.

[47] B. Noble, G.E. Thompson, Precipitation characteristics of Aluminium-Lithium alloys, Met. Sci. J. 5 (1971) 114–120.

[48] X. Zhang, X. Zhou, T. Hashimoto, B. Liu, C. Luo, Z. Sun, Z. Tang, F. Lu, Y. Ma, Corrosion behaviour of 2A97-T6 Al-Cu-Li alloy: The influence of non-uniform precipitation, Corros. Sci. 132 (2018) 1–8.

[49] P. Lequeu, K.P. Smith, A. Daniélou, Aluminum-copper-lithium alloy 2050 developed for medium to thick plate, J. Mater. Eng. Perform. 19 (2010) 841–847.

[50] C.P. Fan, Z.Q. Zheng, M. Jia, J.F. Zhong, B. Cheng, Microstructure and Mechanical Properties of Al-Li Alloy 2397-T87 Rolled Plate, in: Mater. Sci. Forum, Trans Tech Publ, 2014: pp. 249–257.

[51] V.N. Ananiev, The influence of double-ageing on the short-transverse elongation of Al-Li-Cu-Zr rolled plate, in: Mater. Sci. Forum, Aedermannsdorf, Switzerland: Trans Tech Publications, 1996: pp. 865–870.

[52] R. Crooks, Z. Wang, V.I. Levit, R.N. Shenoy, Microtexture, micro structure and plastic anisotropy of AA2195, Mater. Sci. Eng. A. 257 (1998) 145–152.

[53] N.E. Prasad, G. Malakondaiah, Anisotropy of mechanical properties in quaternary Al-Li-Cu-Mg alloys, Bull. Mater. Sci. 15 (1992) 297–310.

[54] A.K. Vasudevan, M.A. Przystupa, W.G. Fricke Jr, Texture-microstructure effects in yield strength anisotropy of 2090 sheet alloy, Scr. Metall. Mater. 24 (1990) 1429–1434.

[55] I. Gutierrez-Urrutia, M.A. Munoz-Morris, D.G. Morris, Recovery of deformation substructure and coarsening of particles on annealing severely plastically deformed Al–Mg–Si alloy and analysis of strengthening mechanisms, J. Mater.

Res. 21 (2006) 329–342.

[56] T. Dorin, F. De Geuser, W. Lefebvre, C. Sigli, A. Deschamps, Strengthening mechanisms of T1 precipitates and their influence on the plasticity of an Al–Cu–Li alloy, Mater. Sci. Eng. A. 605 (2014) 119–126.

[57] B.I. Rodgers, P.B. Prangnell, Quantification of the influence of increased pre-stretching on microstructure-strength relationships in the Al–Cu–Li alloy AA2195, Acta Mater. 108 (2016) 55–67.

[58] J.F. Nie, B.C. Muddle, Microstructural design of high-strength aluminum alloys, J. Phase Equilibria. 19 (1998) 543–551.

**Supporting information**

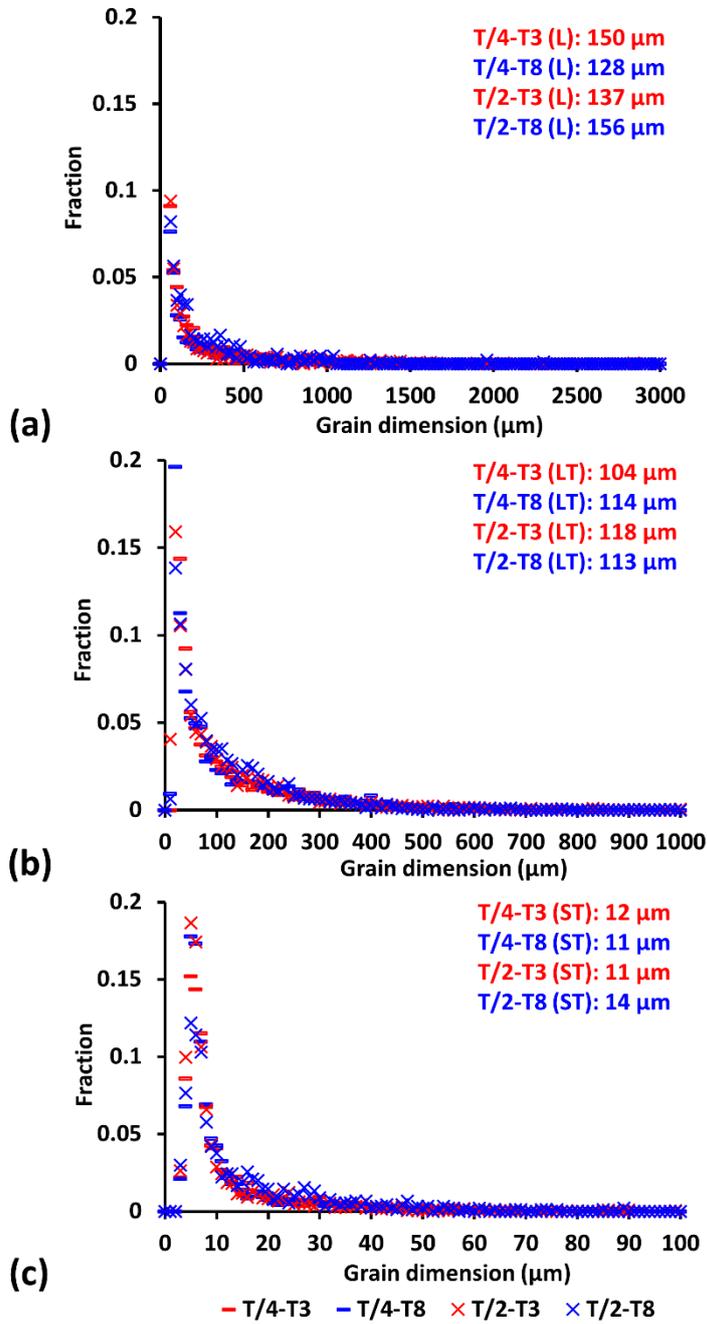

**Figure S1.** The histograms comparing the distribution of grain size along (a) the L, (b) the LT and (c) the ST directions between the T/4 and the T/2 positions for T3 and T8 tempers. The numbers indicate the average size of grains.